\documentclass[a4paper]{article}

\usepackage{INTERSPEECH2022}
\usepackage{multirow}
\usepackage{makecell}
\usepackage{cite}
\usepackage{url}

\title{Dynamic Latency for CTC-Based Streaming Automatic Speech \\Recognition With Emformer}
\name{Jingyu Sun$^{1,2}$, Guiping Zhong$^{1}$, Dinghao Zhou$^{1}$, Baoxiang Li$^{1}$}
\address{
  $^1$SenseTime Research\\
  $^2$Beijing University of Posts and Telecommunications}
\email{\{sunjingyu,zhongguiping,zhoudinghao,libaoxiang\}@sensetime.com}

\begin{document}
\maketitle

\begin{abstract}
An inferior performance of the streaming automatic speech recognition models versus non-streaming model is frequently seen due to the absence of future context.  In order to improve the performance of the streaming model and reduce the computational complexity,  a frame-level model using efficient augment memory transformer block and dynamic latency training method is employed for streaming automatic speech recognition in this paper. The long-range history context is stored into the augment memory bank as a complement to the limited history context used in the encoder. Key and value are cached by a cache mechanism and reused for next chunk to reduce computation. Afterwards, a dynamic latency training method is proposed to obtain better performance and support low and high latency inference simultaneously. Our experiments are conducted on benchmark 960h LibriSpeech data set. With an average latency of 640ms, our model achieves a relative WER reduction of 6.0\% on test-clean and 3.0\% on test-other versus the truncate chunk-wise Transformer.
\end{abstract}

\section{Introduction}
\label{introduction}
Conventional hybrid models have been widely adopted in Automatic Speech Recognition and mostly consist of the acoustic model (AM), pronunciation model (PM) and language model (LM) which are trained separately on different objective and data sets. Compared with hybrid models, end-to-end (E2E) models showed comparable performance by integrating acoustic model, pronunciation model and language model into a single model to achieve joint optimization. Accordingly, many state-of-art models, e.g., connectionist temporal classification (CTC)~\cite{dahl2011context,mohri2008speech,graves2006connectionist}, attention-based encoder-decoder model (AED)~\cite{miao2015eesen,chorowski2014end,chorowski2015attention,zhang2017very}, recurrent neutral network transducer (RNN-T)~\cite{graves2012sequence,graves2013speech,rao2017exploring}, Listen, Attend and Spell (LAS)~\cite{chan2015listen} and Transformer~\cite{vaswani2017attention} have been proposed based on the E2E models.

Streaming ASR is an important scenario in online application, which emit the token as soon as possible after a partial utterance from the speaker. However, the insufficient future context may result in performance degradation. There is a trade-off between latency and accuracy. The AED model and Transformer model for ASR are not feasible for streaming automatic speech recognition, as the global attention mechanism requires all input feature sequence for the calculation of monotonic attention alignment to generate context information. Several methods have been proposed for streaming automatic speech recognition. 

LAS obtains excellent performance, which uses pyramidal BiLSTM and attention mechanism~\cite{chorowski2014end} architecture. Nonetheless, streaming scenarios prevent the direct use of LAS. To facilitate the use of the LAS model for streaming, HMA~\cite{aharoni2016morphological} is put forward to the direct modeling of  monotonic alignment to perform hard attention. Monotonic of chunk attention (MoChA)~\cite{chiu2017monotonic} introduces a novel attention mechanism that retains the benefits of hard monotonic while allowing soft alignments. sMoChA~\cite{miao2019online} is proposed to stable the training of MoChA to obviate exponential decay. ~\cite{inaguma2020minimum} proposes minimum latency training strategies under the MoChA framework. 

To facilitate the use of the transformer-based model for streaming, time-restricted self-attention~\cite{povey2018time} uses a limited length of look-ahead frames for each layer to improve the performance under low latency. However, this method is subject to context leaking in the event of excessive foresight frame lengths such that the foresight context grows linearly with the number of transformer layers. To attenuate context leaking, Transformer-XL~\cite{dai2019transformer} proposes a chunk-wise method in which all frames are visible to other frames in a chunk and caches the previous chunk to reduce the computational complexity. Augment memory transformer~\cite{wu2020streaming,yeh2021streaming} is presented to adopt an augment memory bank to capture long-range history context and reduce computational complexity, which has demonstrated comparable performance.~\cite{cao2021improving} proposes to use self-training to improve the performance of chunk-wise streaming model. However, the performance degradation of the streaming automatic speech recognition models also exist due to the absence of future context under low latency.

In this paper, we solve these drawbacks with a frame-level streaming automatic speech recognition architecture in which Emformer is an encoder to enhance the long-range history context. Emformer~\cite{shi2021emformer}, which is brought forward to reduce computational complexity and applies parallelized block processing training method. To improve the model's performance, we propose a dynamic latency training method. The experiments are conducted on LibriSpeech corpus~\cite{panayotov2015librispeech}. Compared with truncate chunk-wise transformer-based baseline, our model gets a relative WER reduction of 6.0\% on test-clean and 3.0\% on test-other. Moreover, the dynamic latency method gets relative a WER reduction of 5.5\% on test-clean and 1.0\% on test-other versus the efficient augment memory transformer-based model.

The paper is organized as follows: Section \ref{rnnt} gives a brief introduction about CTC criterion, Transformer, Emoformer and proposed dynamic latency training, followed by  experiments and discussions in Section \ref{src_exp}.

\section{Methods}
\label{rnnt}
In this section, we first provide a brief description of CTC and Transformer, then illustrate how to utilize Emformer and dynamic latency training to improve performance.
\subsection{Connectionist Temporal Classification}
The CTC criterion is proposed to map the arbitrary input sequence $X$$=$$(x_1,…,x_t )$ of length $T$ into output sequence $Y$$=$$(y_1,…,y_u )$ of length $U$ directly. In ASR, the length of label $U$ is mostly far smaller than length of speech frames $T$. CTC introduces a blank symbol representing meaningless symbol or silence to map the speech frames into label sequence, resulting in an output label sequence with the same length as the input speech frames $T$. There is a set of labels of $L$ and all labels in CTC are $L^{\prime}=L \mathop{\cup} \{blank\}$. $y_k^t$ is the output probability of unit $k$ at time of $t$, $\pi$ is both the output sequence of length $T$ and an alignment path of label sequence $Y$. The conditional probability of $\pi$ is implicit in Eq.~\ref{eq:alignment} with the conditional independent assumption that the output for different input frame is independent.
\begin{equation}
\label{eq:alignment}
p(\pi \mid X)=\prod_{t=1}^{T} y_{\pi_{t}}^{t}, \pi_{t} \in \mathrm{L}
\end{equation}
CTC criterion defines $\mathcal{B}(\widetilde{Y})$ to map the output sequence with length $T$ into ground truth label sequence $Y$ with length $U$ by removing repeated labels and blank symbols from $\widetilde{Y}$, where $\widetilde{Y}$ is an alignment path of the ground truth $Y$. CTC calculates the sum of the probability of all alignment paths for label sequence $Y$:
\begin{equation}
p(Y \mid X)=\sum_{\pi \in \mathcal{B}^{-1}(Y)} p(\pi \mid X)
\end{equation}
Then CTC objective function $\mathcal{L}$ is to optimize the negative log probability of all alignment paths for label sequence $Y$:
\begin{equation}
    \mathcal{L}(X)=-\operatorname{log}p(Y \mid X)
\end{equation}
Direct calculation of the probability of all possible alignment paths is laborious due to the computational complexity and the forward-backward algorithm is introduced~\cite{yu2003efficient} to calculate the probability of the given label sequence.
\subsection{Transformer}

Transformer is proposed in~\cite{vaswani2017attention} and is an attention-based encoder-decoder model. 
The Transformer model is a stack of transformer layers that consist of the  self-attention network (SAN), feed-forward network (FFN), layernorm and residual connection. The dot-product attention mechanism in SAN can be represented below:
\begin{equation}
 { Attn }(Q, K, V)= \operatorname{Softmax}\left(\frac{Q K^{T}}{\sqrt{d_{k}}}\right)V
\end{equation}
Where $d_{k}$ represents the dimension of $K$, $\frac{1}{\sqrt{d_{k}}}$ is a scaling factor and $Q$, $K$, and $V$ refer to the query, key and value embedding, which are projected to high-level linear vector space using learnable matrices $W_{q}$,$W_{k}$ and $W_{v}$:
\begin{equation}
    Q = XW_{q},K = XW_{k},V = XW_{v}
\end{equation}
Subsequently, a feed-forward network is applied, consisting of two linear fully-connected layers follow by a ReLU activation function, as presented in Eq.~\ref{eq:ffn}:
\begin{equation}
\label{eq:ffn}
    \operatorname{FFN}(x)=\operatorname{ReLU}(xW_1+b_1)W_2+b_2
\end{equation}
Another beneficial multi-head attention (MHA) mechanism is also put forward in~\cite{vaswani2017attention}, which projects query, key and value embedding to several separately linear vector subspace and applies attention function in parallel. Then all the embedding are concatenated and projected together with learnable matrix $W_{o}$:
\begin{equation}
\text { $head_{i}$ }=\operatorname{Attn}(W_{k}^{i}Q_{i},W_{k}^{i}K_{i},W_{k}^{i}V_{i})
\end{equation}
\begin{equation}
\text { MHA }(Q, K, V)=\operatorname{concat}({head_{1},..., head_{h}})W_{o}
\end{equation}
where $h$ is the number of head, $W_{q}^{i}$,$W_{k}^{i}$ and $W_{v}^{i}$ 
are learnable matrices for i-th attention head.
\subsection{Architecture of Proposed Model}
\subsubsection{Emformer}
\begin{figure}[t]
	\centering
	\includegraphics[width=\linewidth]{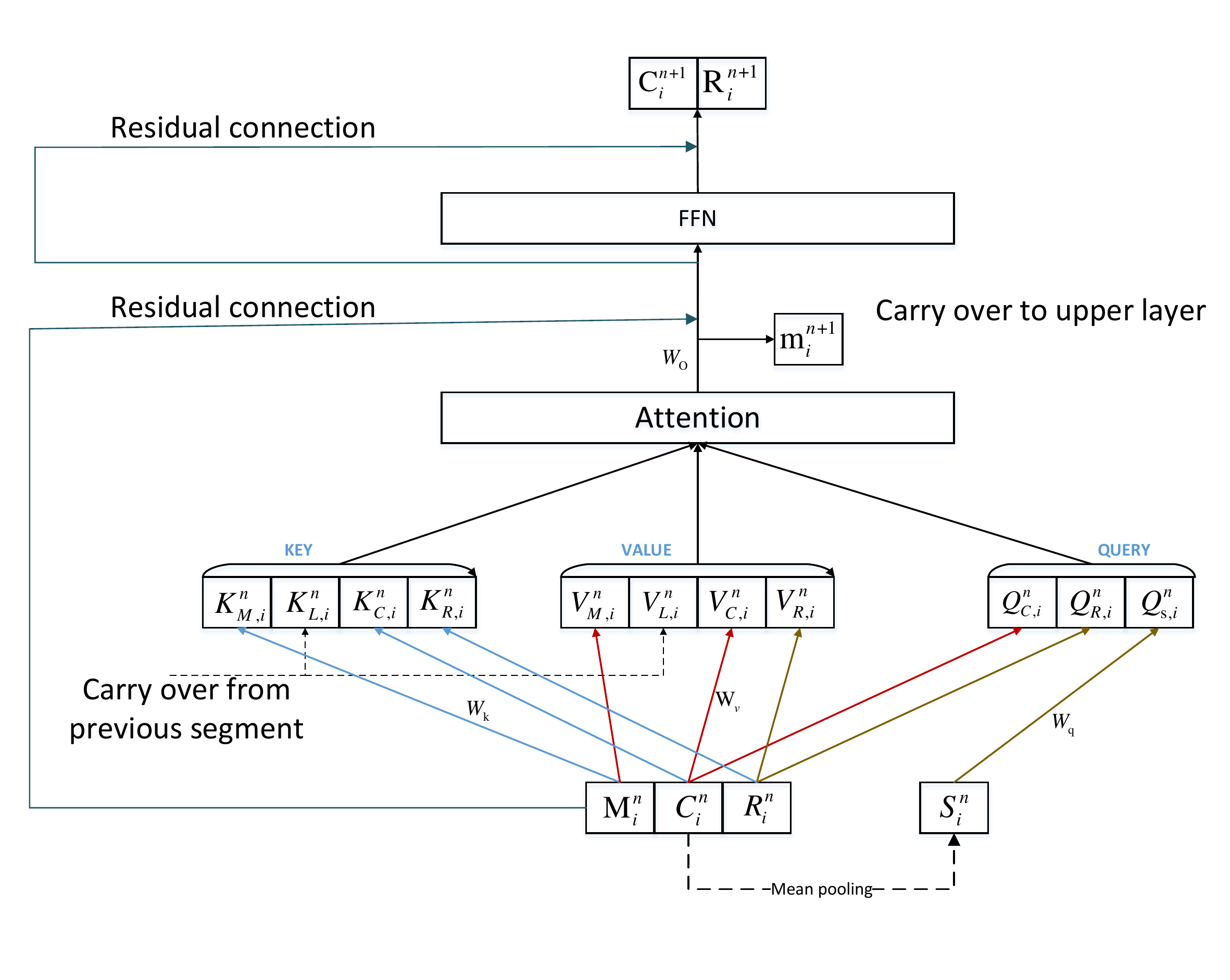}
	\caption{The architecture of Emformer encoder, summary context is extracted from center context as first memory block}
	\label{fig:DL}
\end{figure}

The computational complexity of the Transformer is quadratic growth with the length of input speech sequence, which could be resolved partly by using limited history context to reduce the computation of attention mechanism, but such method may lead to performance degradation. Emformer uses augmented memory bank~\cite{wu2020streaming} to carry out long-range history context to augment history context under limited history context in the encoder. Emformer first splits the input sequence into blocks and all $C$ input frames are calculated in one go with assumption of each block containing $C$ input frames. 

The forward step using augmented memory transformer in Emformer is illustrated in Figure \ref{fig:DL} and the i-th block's attention output ${X}_{i}^{n+1}$ is calculated as follow:
\begin{equation}
\begin{gathered}
{K}_{i}^{n}=\left[{W}_{\mathrm{k}} {M}_{i}^{n}, {K}_{L, i}^{n}, {W}_{\mathrm{k}} {C}_{i}^{n}, {W}_{\mathrm{k}} {R}_{i}^{n}\right], \\
\end{gathered}
\end{equation}
\begin{equation}
{V}_{i}^{n}=\left[{W}_{{v}} {M}_{i}^{n}, {V}_{L, i}^{n}, {W}_{{v}} {C}_{i}^{n}, {W}_{{v}} {R}_{i}^{n}\right], \\
\end{equation}
\begin{equation}
{\left[\widetilde{{C}}_{i}^{n}, \widetilde{{R}}_{i}^{n}\right]=\operatorname{LayerNorm}([ \boldsymbol{C}_{i}^{n}, \boldsymbol{R}_{i}^{n}])}, \\
\end{equation}
\begin{equation}
{Z}_{\mathrm{C}, i}^{n}=\operatorname{Attn}\left({W}_{\mathrm{q}} \widetilde{{C}}_{i}^{n}, {K}_{i}^{n}, {V}_{i}^{n}\right)+{C}_{i}^{n}, \\
\end{equation}
\begin{equation}
{Z}_{\mathrm{R}, i}^{n}=\operatorname{Attn}\left({W}_{{q}} \widetilde{{R}}_{i}^{n}, {K}_{i}^{n}, {V}_{i}^{n}\right)+{R}_{i}^{n}, \\
\end{equation}
\begin{equation}
{m}_{i}^{n}=\operatorname{Attn}\left({W}_{\mathrm{q}} {s}_{i}^{n} ; {K}_{i}^{n}, {V}_{i}^{n}\right), \\
\end{equation}
\begin{equation}
{M}_{i}^{n+1}=\left[{M}_{i}^{n+1}, {m}_{i}^{n}\right], \\
\end{equation}
\begin{equation}
\widetilde{{X}}_{i+1}^{n}=\text { FFN }\left(\operatorname{LayerNorm}\left(\left[{Z}_{\mathrm{C}, i}^{n}, {Z}_{\mathrm{R}, i}^{n}\right]\right)\right), \\
\end{equation}
\begin{equation}
\begin{gathered}
{X}_{i}^{n+1}=\text { LayerNorm }\left(\widetilde{{X}}_{i}^{n+1}+\left[{Z}_{\mathrm{C}, i}^{n}, {Z}_{\mathrm{R}, i}^{n}\right]\right)
\end{gathered}
\end{equation}

Where $K_i^n$ and $V_i^n$ represent key and value embedding that concatenates the left context $L_i^n$, future block $R_i^n$ and $C_i^n$. Memory bank $M_i^n$ is concatenated with current center block $C_i^n$ for the i-th block. The $K_{L,i}^n$ and $V_{L,i}^n$ refer to key and value cached from the previous block, note that the center block $C_{i-1}^n$ for (i-1)-th block is left context to $C_i^n$, for the reason that the overlapping of $C_{i-1}^n$ with $L_i^n$ leads to entail recomputation for every block, Emformer caches the projections for left context to avoid additional computation.

The query in Emformer block can be written as ${Q}_{i}^{n}=\left[ {W}_{\mathrm{q}} {C}_{i}^{n}, {W}_{\mathrm{q}} {R}_{i}^{n},{W}_{\mathrm{q}} {q}_{i}^{n} \right] $, which is  easier for understanding by deviding them apart, where $s_i^n$ refers to summarization vector, $s_i^n$ is only applied before attention function, $[Z_{C,i}^n,Z_{R,i}^n ]$ then is applied to the residual connection and FFN. The attention output $X_i^{n+1}$ is fed to the next layer.

\textbf{Carry Memory Vector From Previous Layer}.\ \ 
$s_i^n$ is computed from the center block by average pooling and used to generate memory vector for current center block $m_i^n$,   $M_i^n$$=$$(m_1^{n-1},…,m_{i-1}^{n-1})$, which is the augmented memory bank storing every memory vector from the previous layer. Models can be trained in a parallel way to boost training and better manage the utilization of GPU computation resources.

\textbf{Disallow Attention Between Summary Vector With Memory Bank}.\ \ 
By assigning the attention weight between summary vector and memory vector to zero, Emformer can stabilize the training and improves the performance for long speech sequence.

\textbf{Alleviate Look-ahead Context Leaking}.\ \ 
Attention mask~\cite{povey2018time,zhang2020transformer} is applied in Emformer to restrict the reception field in each layer for every block to enable training in a parallel manner and avoid look-ahead context leaking.
\subsubsection{Dynamic Latency Training}

\begin{figure}[t]
	\centering
	\includegraphics[width=\linewidth]{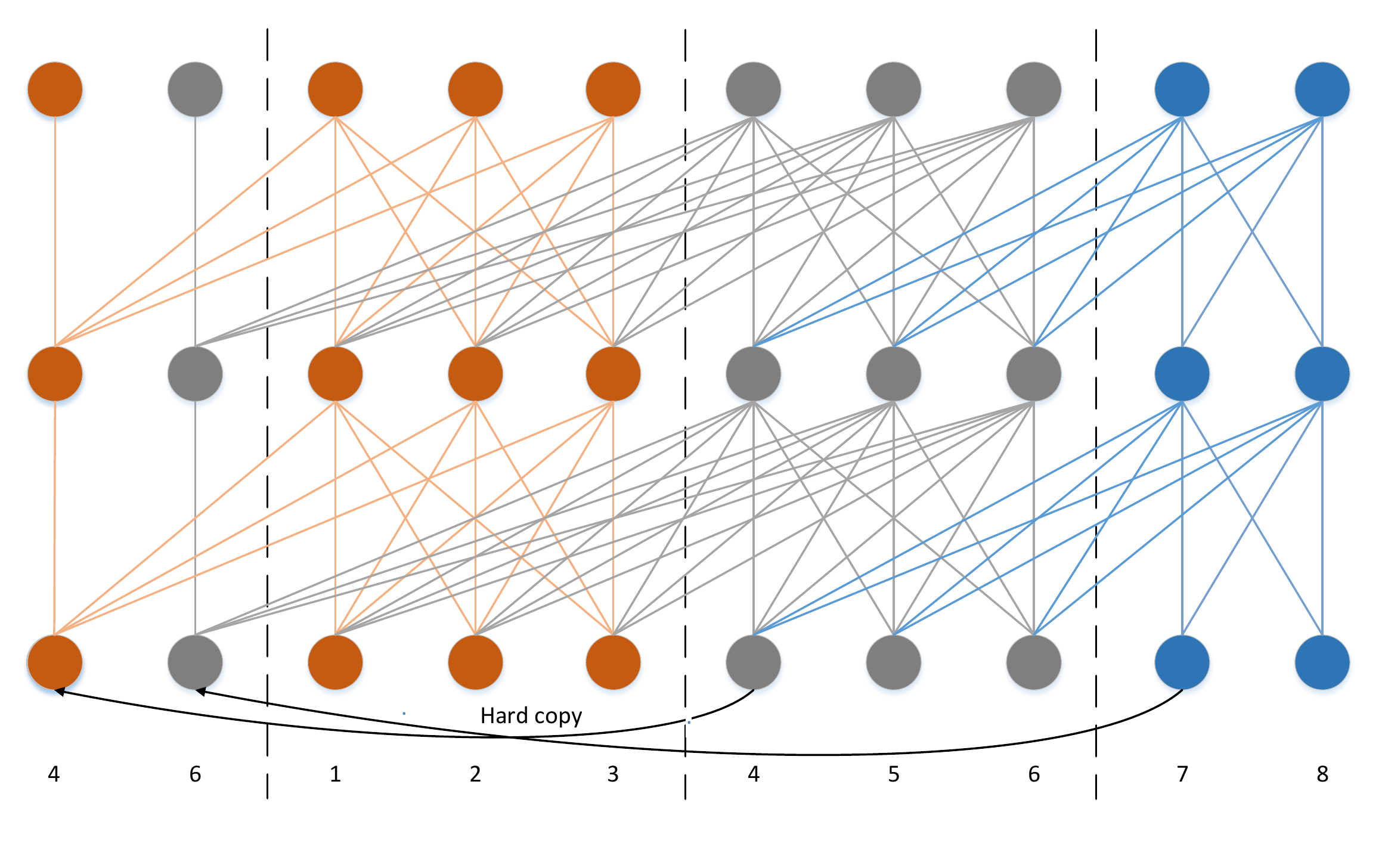}
	\caption{ Illustration of hard copy to avoid look-ahead context leaking. It's an example of model with center context size 3 and future context size 1. Frame 1,2,3,4 are visible to frame 3 and future context 4 is copied to the beginning of the sequence so that only the frame of the left side is utilized to calculate the attention for current block.}
	\label{fig:hard_copy}
\end{figure}
This method is utilized to augment the future context information without increasing additional latency. In dynamic latency training, we maintain the left context of each layer unchange to alleviate quadratic growth of computational complexity and sample future context length from a given discrete distribution. All future context with the sampled length for each block is copied to the beginning of the input feature frames for each batch data as illustrated in Figure~\ref{fig:hard_copy}. All models' EIL is set to 640ms for inference. In the training stage, future context is sampled from a list of future context configurations. For example, we set $[0ms,320ms,640ms]$ for a block size with 640ms and set $[0ms,160ms,320ms]$ for a block size with 960ms.

In our experiments, an overwhelming volume of future contexts in configuration may lead to performance degradation. We limit the number of future context length configurations sampled from various latency to avoid performance degradation. Compared with a small future context length, a large future context length is associated with better performance and higher latency. The parameters of the model are shared for different future context configuration and we optimize both the configurations with small future context and large future context, and the configuration with small future context gains better performance versus with constant future context length. This method enlarges the receptive field of the center block and forces the model to learn more future information for the center block.

\section{Experiments}
\label{src_exp}
\subsection{Experimental Setup}
The experiments are conducted on LibriSpeech corpus~\cite{panayotov2015librispeech}, which contains 960 hours of book reading audios. There are two subsets of development data and evaluation data in LibriSpeech. The test-clean and dev-clean contain simple and clean audios, test-other and dev-other contain complex and noisy audios. The model is evaluated on test-clean and test-other subsets. All models utilize feature extraction followed by Spec-Augment~\cite{park2019specaugment}.
The feature embedding fed into the encoder is an 80-dimensional log filter-bank feature with a 10ms frame shift, 25ms window size and stacked by 4 frames. Following~\cite{zhang2017very}, two VGG blocks~\cite{wang2020transformer} are introduced before the transformer layers. Each VGG block consists of two consecutive 3-by-3 convolution layers, ReLU~\cite{agarap2018deep} activation function and pooling layer. The first VGG block has 32 channels and the second VGG block has 64 channels, 2-by-2 max-pooling is used in each block with 2 stride frames.

The encoder has 12 transformer layers, each layer has SAN module with 8 attention heads, model dimension 768 and FFN with dimension 2048. Dropout~\cite{srivastava2014dropout} is set to 0.1 for all transformer layers. CTC criterion is applied to optimize the CTC loss. All left context is set at 2560ms to reduce the computational complexity.
Models are trained with the Adam~\cite{kingma2014adam} optimizer for 400000 steps, warming-up stage for 20000 steps, maintained constant for 100000 steps and then linearly decay for the remainder with the peak learning rate set to $1e^{-4}$. 
In inference, the 4-gram language model trained with LibriSpeech training transcript is used in beam-search set to 100. All of experiments are conducted on 8 NVIDIA Tesla V100 GPUs.

\subsection{Algorithmic Latency Induced By Encoder (EIL)}
In decoding, the latency originates from the waiting time with respect to center block size and the future context size. Latency for the most left frame in the center block is center block size plus future context size. Latency for the rightest frame in the center block is future context size. We use algorithmic latency induced by the encoder (EIL) to calculate the average latency of all frames in the center block:
\begin{equation}
    EIL=0.5\times block_{size} + future_{size}
\end{equation}

\subsection{Performance of Proposed Method}
As illustrated in Table \ref{640_baseline}, we evaluate the performance of Emformer and truncate chunk-wise transformer with latency 640ms. The truncate chunk-wise Transformer baseline is trained on LibriSpeech and achieves a WER of 4.04/11.54 on test-clean/test-other. Emformer with the memory size set to 4 achieves a WER of 4.02/11.41. Compared with truncate chunk-wise Transformer, Emformer gets a relative WER reduction of 0.5\% on test-clean and 1.2\% on test-other, which shows that the memory bank for storing long-range history context to augment history information is slightly beneficial for the center block.

\begin{table}[th]
	\caption{Comparison of Emformer, Transformer for center context 640ms and other results. The memory size for Emformer is set at 4 constant for all experiments. Center context and future context are set to 640ms and 320ms respectively. S1 and S2 are results in~\cite{cao2021improving} with similar latency and infinite history context. S1 indicates Time-restricted Transformer, S2 indicates Chunk-wise Transformer }
	\label{640_baseline}
	\centering
	
	\setlength{\tabcolsep}{0.7mm}{
    \begin{tabular}{lcccccc}
    		\toprule
    		model & \makecell[c]{left\\context}&\makecell[c]{Center \\ Context} & \makecell[c]{Future \\ Context} & \makecell[c]{Memory \\ Size} & \makecell[c]{test\\clean} & \makecell[c]{test\\other} \\
    		\midrule
    		S1 &inf &- &480ms &- &4.4 &13.0\\
            S2 &inf &960ms &0 &-  &3.9 &11.4\\
    		\midrule
    	    Transformer &2560ms & 640ms & 320ms & - & 4.04 & 11.54 \\
    		Emformer &2560ms &640ms & 320ms & 4 & 4.02 & 11.41 \\
    		\bottomrule
	\end{tabular}
	}
\end{table}

While Table~\ref{960_baseline} shows the results when setting the center block to 960ms and EIL with 640ms. Transformer achieves a WER of 3.98/11.81 on test-clean/test-other and Emformer achieves a WER of 3.98/11.58. Compared with the truncate chunk-wise transformer with the same latency, Emformer gets a relative WER reduction of 2\% on test-other. Emformer performs slightly better on test-other demonstrating the benefit of the memory bank.

\begin{table}[th]
	\caption{Comparison of Emformer and Transformer for center context 960ms. Center context and future context are set to 960ms and 160ms respectively.}
	\label{960_baseline}
	\centering
	\setlength{\tabcolsep}{1.6mm}{
    \begin{tabular}{lccccc}
    		\toprule
    		model & \makecell[c]{Center \\ Context} & \makecell[c]{Future \\ Context} & \makecell[c]{Memory \\ Size} & \makecell[c]{test\\clean} & \makecell[c]{test\\other} \\
    		\midrule
    	    Transformer & 960ms & 160ms & - & 3.98 & 11.81 \\
    		\midrule
    		Emformer  & 960ms & 160ms & 4 & 3.98 & 11.58 \\
    		\bottomrule
	\end{tabular}
	}
\end{table}

As illustrated in Table\ref{640_exp} and Table\ref{960_exp}, the performance of dynamic latency training with different context length configurations is evaluated. In Table \ref{640_exp}, Emformer trained with $E3$ that uses $[0ms,320ms,1280ms]$ as future context list, achieves the best performance with a WER of 3.80/11.37 and gets a relative WER reduction of 6.0\% on test-clean and 1.5\% on test-other. Compared with a WER of 3.80/11.53 achieved by $E2$, higher latency configuration gets more future context and outperforms low latency in inference. $E1$ shows that a close gap between each context length may cause model suffering from dynamic latency training method.
\begin{table}[th]
	\caption{WER(\%) of Emformer using dynamic latency training with different future context length configuration under center context 640ms and $EIL$ 640ms. E1 indicates Emformer with a future configuration of $[320ms,400ms,480ms,560ms,640ms]$, E2 indicates Emformer with a future configuration of $[0ms,320ms,640ms]$ and E3 indicates Emformer with a future configuration of $[0ms,320ms,1280ms]$.}
	\label{640_exp}
	\centering
	
	\setlength{\tabcolsep}{2.9mm}{
    \begin{tabular}{lcccc}
    		\toprule
    		model & \makecell[c]{Center \\ Context} & \makecell[c]{Future \\ Context} & \makecell[c]{test\\clean} & \makecell[c]{test\\other} \\
    		\midrule
    	    Transformer & 640ms & 320ms & 4.04 & 11.54 \\
    		\midrule
    		\qquad+E1 & 640ms & 320ms & 3.86 & 11.52 \\
            \qquad+E2 & 640ms & 320ms & 3.80 & 11.53 \\
            \qquad+E3 & 640ms & 320ms & \textbf{3.80} & \textbf{11.37} \\
    		\bottomrule
	\end{tabular}
	}
\end{table}

\begin{table}[th]
	\caption{WER(\%) of Emformer using dynamic latency training with different future context length configuration under center context 960ms and $EIL$ 640ms. E4 indicates Emformer with a future configuration of $[160ms,1280ms]$, E5 indicates Emformer with a future configuration of $[0ms,160ms,320ms]$ and E6 indicates Emformer with a future configuration of $[0ms,160ms,640ms]$.}
	\label{960_exp}
	\centering
	\setlength{\tabcolsep}{3.0mm}{
    \begin{tabular}{lcccc}
    		\toprule
    		model & \makecell[c]{Center \\ Context} & \makecell[c]{Future \\ Context} & \makecell[c]{test\\clean} & \makecell[c]{test\\other} \\
    		\midrule
    	    Transformer & 960ms & 160ms & 3.98 & 11.81 \\
    		\midrule
    		\qquad+E4 & 960ms & 160ms & 3.87 & 11.66 \\
            \qquad+E5 & 960ms & 160ms & 3.87 & 11.56 \\
            \qquad+E6 & 960ms & 160ms & \textbf{3.80} & \textbf{11.46} \\
    		\bottomrule
	\end{tabular}
	}
\end{table}
In Table \ref{960_exp}, all models' center block size is set to 960ms. Emformer trained with $E6$ gains the best performance with a WER of 3.80/11.46 and gets a relative WER reduction of 4.6\% on test-clean and 3.0\% on test-other. Comparing $E6$ and $E5$ with a WER of 3.87/11.56, a higher latency configuration obtains more future context and outperforms low latency in inference, which is also proved in Table~\ref{640_exp}.

As illustrated both in Table~\ref{640_exp} and Table~\ref{960_exp}, the proposed model's center block size is constant while future context size is various and the size is sampled from a configuration list. The best models with dynamic latency training have future context length of 0ms in a list, which indicates that the future context configuration including both higher future context and lower future context can improve the performance of feature extracting in block and acquire more context from training.

\section{Conclusions}
\label{sec_con}
In this paper, we aim to improve the performance of the streaming model and reduce the computational complexity under low latency. Firstly, we explore the performance of Emformer using memory size 4 with block size 640ms and 960ms. Furthermore, to improve the performance, we propose a dynamic latency training method to augment the future context for streaming automatic speech recognition. We improve the performance of truncate chunk-wise Transformer using both Emformer for augmenting the long-range history context and dynamic latency training method. Experiments show that our proposed method achieves a relative WER reduction of 6.0\% on test-clean and 1.5\% on test-other under center block size 640ms, achieves relative WER reduction 4.6\% on test-clean and 3.0\% on test-other under center block size 960ms.

\bibliographystyle{IEEEtran}

\bibliography{template}

\end{document}